\documentclass{article}
\usepackage{cite,spconf,amsmath,graphicx,enumitem,amssymb,color,bm}
\usepackage{algorithmic}
\usepackage{graphicx}
\usepackage{textcomp}
\usepackage{xcolor}
\usepackage[caption=false, font=footnotesize, labelfont=rm, textfont=rm]{subfig}
\usepackage{epstopdf}
\usepackage{mathrsfs}
\title{Wireless Deep Speech Semantic Transmission}

\name{Zixuan~Xiao $^{\star}$,~Shengshi~Yao$^{\star}$,~Jincheng~Dai$^{\star}$,~Sixian~Wang$^{\star}$,~Kai~Niu$^{\star \dag}$,~Ping~Zhang$^{\star}$
\thanks{This work was supported in part by the National Natural Science Foundation of China under Grant 92067202, Grant 62001049, and Grant 62071058, in part by the Beijing Natural Science Foundation under Grant 4222012. \emph{(Corresponding author: Jincheng Dai)}}
}

\address{
\normalsize $^{\star}$ Beijing University of Posts and Telecommunications, Beijing, 100876, China\\
\normalsize $^{\dag}$ Peng Cheng Laboratory, Shenzhen, 518067, China\\
\normalsize Email: daijincheng@bupt.edu.cn
}

\begin{document}
\ninept
\maketitle
\begin{abstract}
In this paper, we propose a new class of high-efficiency semantic coded transmission methods for end-to-end speech transmission over wireless channels. We name the whole system as \emph{deep speech semantic transmission (DSST)}. Specifically, we introduce a nonlinear transform to map the speech source to semantic latent space and feed semantic features into source-channel encoder to generate the channel-input sequence. Guided by the variational modeling idea, we build an entropy model on the latent space to estimate the importance diversity among semantic feature embeddings. Accordingly, these semantic features of different importance can be allocated with different coding rates reasonably, which maximizes the system coding gain. Furthermore, we introduce a channel signal-to-noise ratio (SNR) adaptation mechanism such that a single model can be applied over various channel states. The end-to-end optimization of our model leads to a flexible rate-distortion (RD) trade-off, supporting versatile wireless speech semantic transmission. Experimental results verify that our DSST system clearly outperforms current engineered speech transmission systems on both objective and subjective metrics. Compared with existing neural speech semantic transmission methods, our model saves up to 75\% of channel bandwidth costs when achieving the same quality. An intuitive comparison of audio demos can be found at \emph{https://ximoo123.github.io/DSST}.
\end{abstract}
%
%\vspace{-0.2cm}
\section{Introduction}
\label{sec:introduction}
%\vspace{-0.2cm}

%Source coding and channel coding are two essential steps in modern data transmission\cite{shannon1948mathematical}. Source coding is in charge of compressing speech signals into bit sequences, and channel coding fortified by error correcting codes, provides error-free transmission in the channel. However, the limits of the separation-based design begin to emerge with more requirements and challenge on speech transmission. On the one hand, traditional separated coding schemes suffer from error propagation. When the channel decoding process exists mistakes, the reconstructed speech quality shows substantial degradation, although the source coding shows great compression performance. On the other hand, cliff effect of the end-to-end performance exist in separation-based schemes when the channel quality deteriorates below the minimum channel quality to allow successful decoding. In other words, the speech becomes un-decodable due to the channel quality falling below what the channel code anticipates.

%However, with the wide deployment of Internet of Things applications, these semantic-irrelative schemes are no longer ideal as they transmit bit sequence, which contains information not relevant to the receiver task.
%Hence, it is very time to set a semantic communication system to boost end-to-end communications performance

State-of-the-art (SOTA) engineered speech transmission methods over the wireless channel can usually be separated into two steps: source coding and channel coding \cite{shannon1948mathematical}. Source coding performs a linear transform on the waveform, which removes redundancy, and channel coding provides error-correction against imperfect wireless channels. Nevertheless, the optimality of separation-based schemes holds only under infinite coding blocklengths and unlimited complexity, which is impractical for engineered communication systems. Moreover, a severe ``\emph{cliff effect}'' exists in separation-based schemes where the performance breaks down when the channel capacity goes below the communication rate. The limits of this separation-based design begin to emerge with more demands on low-latency wireless speech transmission applications.

To tackle this, it is very time to bridge source and channel coding to boost end-to-end communications performance. Recent advances in deep learning have led to increased interest in solving this problem that employs the nonlinear property and the end-to-end learning capability of neural networks \cite{kurka2021bandwidth,farsad2018deep,Dai,Dai9852388}. The whole system is referred to as \emph{semantic coded transmission (SCT)}. In this paper, we focus on SCT for speech sources. Previous works in this topic, e.g., DeepSC-S \cite{weng2021semantic}, have shown the potential of SCT to avoid cliff effect, which however ignore the tradeoff between the channel bandwidth cost and the end-to-end transmission quality. In this paper, we inherit the fundamental principle of classical speech coding but integrates the nonlinear superiority of neural networks for realizing a class of high-efficiency speech SCT systems, which are named ``\emph{deep speech semantic transmission (DSST)}''.

Specifically, in the time domain, we sample the source speech waveform and feed them into a neural network based \emph{semantic analysis transform} to produce latent features, which are indeed a representation of the source speech. These latent feature embeddings are further fed into a neural \emph{joint source-channel encoder} to produce the channel-input sequence. During this process, a critical latent prior is imposed on the latent features, which variationally estimates the entropy distribution on the semantic feature space. It guides the source-channel coding rate allocation to maximize the system coding gain. The whole DSST system is formulated as an optimization problem aiming to minimize the transmission rate-distortion (RD) performance. Our system is versatile: one model can arbitrary tradeoff between rate and distortion. We carry out experiments under the additive white Gaussian noise (AWGN) channel and the practical COST2100 fading channel \cite{liu2012cost}. Results verify that our DSST system clearly outperforms current engineered speech transmission systems on both objective and subjective metrics. Compared with existing neural speech semantic transmission methods, our model saves up to 75\% of channel bandwidth costs when achieving the same quality.

\section{Our Proposed DSST System}

\subsection{Architecture}
\label{2.1}

Our network architecture, shown in Fig. \ref{fig1}, comprises a semantic transform network, a prior network, and a Deep joint source-channel coding (JSCC) network.
Given a speech sample $\bm{x}$ in time domain, it is first transformed into semantic latent representation $\bm{y}$ in the semantic feature domain by a DNN-based nonlinear analysis transform $g_{a}$. Then, the latent feature $\bm{y}$ is fed into both prior encoder $h_{a}$ and Deep JSCC encoder $f_{e}$.
On the one hand, $h_{a}$ captures the temporal dependencies of $\bm{y}$ and utilizes a learnable entropy model to characterize the semantic importance of rate allocation. On the other hand, $f_{e}$ encodes $\bm{y}$ as channel-input sequence $\bm{s}$ for transmission over the wireless channel directly and the received sequence is $\bm{\hat{s}}$. In this paper, we consider AWGN channel and COST2100 wireless fading channel, such that $\hat{\bm{s}}$ can be written as $\bm{\hat{s}} = W(\bm{s|h})=\bm{h} \odot \bm{s}+\bm{n} $, where $\bm{h}$ is the CSI vector, $\odot$ is the element-wise product and $\bm{n}$ denotes the noise vector, it is independently sampled from Gaussian distribution, i.e. $\bm{n} \sim \mathcal{CN}(\bm{0},\sigma_{n}^{2}I)$. The structure of the receiver is with a mirrored architecture. As illustrated in Fig. \ref{fig1}, $f_{d}$ is Deep JSCC decoder which aims to recover latent representation $\hat{\bm{y}}$, $g_{s}$ is semantic synthesis transform for reconstructing source signals $\hat{\bm{x}}$ and $h_{s}$ is prior decoder. Hence, the total link of the DSST system is formulated as follows:
\begin{figure}[t]
\centering %textwidth值小于0.25，或者linewidth小于0.5，不过这里设置textwidth比设置linewidth效果好一些
\includegraphics[scale=0.345]{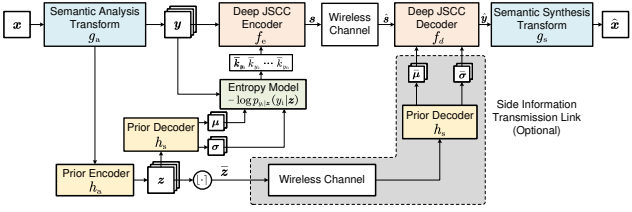}  %0.345
\vspace{-1em}
\caption{The architecture of DSST.}
\vspace{-1.8em}
\label{fig1}
\end{figure}

\vspace{-1.4em}
\begin{equation}\label{eq_speech_trans_process}
\begin{split}
& {{\bm{x}}} \xrightarrow{{{g_a(\cdot)}}} {{\bm{y}}}
\xrightarrow{{{f_e(\cdot)}}} {{\bm{s}}}
\xrightarrow{{{W(\cdot |\bm{h})}}} {{\bm{\hat s}}}
\xrightarrow{{{f_d(\cdot)}}} {{\bm{\hat y}}}
\xrightarrow{{{g_s(\cdot)}}} {{\bm{\hat x}}} \\ &
\text{with the latent prior } {{\bm{y}}} \xrightarrow{{{h_a(\cdot)}}} {{\bm{z}}}
\xrightarrow{{{h_s(\cdot)}}} {{\left\{{\bm{\mu}},{\bm{\sigma}}\right\}}}.
\end{split}
\vspace{-0.8em}
\end{equation}
%\vspace{-0.8em}
Moreover, we design an optional transmission according to the actual channel situation. When the channel capacity is large enough, we transmit the side information $\bm{z}$ to the receiver by reliable digital coding schemes as a hyperprior to acquire higher decoding gain. When there is a shortage of channel capacity, $\bm{z}$ will not be transmitted, and the system performance will be slightly degraded as a result.

\subsection{Variational Modeling and Source-Channel Coding}
\label{2.2}
In our intuition, the speech signal at the silent time almost carries few information. Therefore, signals with zero magnitudes should be allocated less channel bandwidth and vice versa. If we assign the same coding rate for semantic features of all frames, the coding efficiency cannot be maximized. To tackle this problem, we design an entropy model to accurately estimate the distribution of $\bm{y}$ such that it evaluates the importance of different semantic features. Moreover, we try to make the Deep JSCC encoder able to aware of the semantic characteristics of the source and reasonably allocate coding rate according to the semantic importance.

%We utilize prior decoder $h_{s}$ to predict them
%Hence, the entropy of each speech frame is computed as $r_{i} = - \log P_{y_{i}|\bm{z}}$.

Our entropy model is illustrated in the middle part of Fig. \ref{fig1}. Following the work of \cite{balle2018variational,balle2020nonlinear}, the semantic feature $\bm{y}$ is variational modeled as a multivariate Gaussian with the mean $\bm{\mu}$ and the standard deviation $\bm{\sigma}$. Thus the true posterior distribution of $\bm{y}$ can be modeled by a fully factorized density model as:
%\vspace{-0.7em}
\begin{equation}
\begin{split}
& p_{\bm{y | z}}(\bm{y | z}) = \prod_{i} \left(\mathcal{N}(y_{i}| \mu_{i},\sigma_{i}) \ast \mathcal{U}(-\frac{1}{2},\frac{1}{2})\right)(y_{i}) \\&
\text{with } (\bm{\mu},\bm{\sigma})=h_{s}(\bm{z}),
\end{split}
\vspace{-0.8em}
\end{equation}
where $\ast$ is convolutional operation, $\mathcal{U}$ denotes a uniform distribution, it utilized to match the prior to the marginal such that the estimated entropy $-\log p_{\bm{y}|\bm{z}}(\bm{y}|\bm{z})$ is non-negative. During optimizing, the actual distribution $q_{\bm{y | z}}$ created by $h_{a}$ will gradually approximate the true distribution $p_{\bm{y | z}}$, hence the entropy model estimates distribution of $\bm{y}$ accurately. The entropy value of each frame $-\log p_{y_{i}|\bm{z}}(y_{i}|\bm{z})$ will be fed into JSCC encoder and guides the allocation of coding rate accordingly. If the entropy model indicates $y_{i}$ of high entropy, the encoder will allocate more resources to transmit it. The primary link channel bandwidth cost $K_{y}$ for transmitting $\bm{y}$ can be written as:
%\begin{equation}
%K = \sum_{i=1}^{B} \overline{k}_{y_{i}} = \sum_{i=1}^{B} \eta_{y}r_{i}= \sum_{i=1}^{B} Q^{'}(-\eta_{y}logp_{y_{i}|\bm{z}}(y_{i}|\bm{z}))
%\end{equation}
\vspace{-0.5em}
\begin{equation}
K_{y} = \sum_{i=1}^{B} \bar{k}_{y_{i}} = \sum_{i=1}^{B} Q(k_{y_{i}})= \sum_{i=1}^{B} Q(-\eta_{y}\log p_{y_{i}|\bm{z}}(y_{i}|\bm{z})),
\end{equation}
where $B$ is the number of frames of the speech signal, $\eta_{y}$ is a hyperparameter to balance channel bandwidth cost from the estimated entropy, $\bar{k}_{y_{i}}$ is the bandwidth consumption of frame $i$, $Q$ denotes a scalar quantization which contains $n$ quantized values.

\subsection{Optimization Goal}
\begin{figure}[t]
\centering
\vspace{-0.8em}
\includegraphics[scale=0.425]{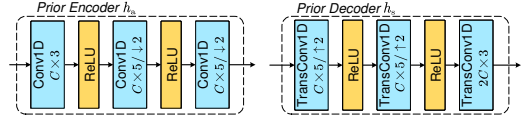}
\vspace{-1.1em}
\caption{The architecture of prior network. Conv1D $C \times k$ is a one-dimension (1D) convolution layer with $C$ channels and $k$ filters. The following $\downarrow 2$ indicates downscaling with a stride of 2.}
\label{fig2}
\vspace{-1.7em}
\end{figure}
The optimization goal of the DSST system is to attain an RD trade-off between channel bandwidth cost and speech reconstruction quality. It is proven by Ball$\mathrm{\acute{e}}$ that it is more reasonable to model the whole model as a variational autoencoder \cite{balle2018variational}, such that minimizing the KL divergence is equivalent to optimizing the model for RD performance. Hence, the goal of inference model is creating a variational density $q_{\bm{\hat{s},\tilde{z}|x}}$ to approximate the true posterior $p_{\bm{\hat{s},\tilde{z}|x}}$. The RD function can be approximated as
\vspace{-0.7em}
\begin{equation}
\begin{split}
& L_{\mathrm{RD}}=\mathbb{E}_{\bm{x} \sim p_{\bm{x}}}D_{\mathrm{KL}}[q_{\bm{\hat{s},\tilde{z}|x}}||p_{\bm{\hat{s},\tilde{z}|x}}] \Leftrightarrow \mathbb{E}_{\bm{x} \sim p_{\bm{x}}}\mathbb{E}_{\bm{\hat{s}},\bm{\tilde{z}} \sim q_{\bm{\hat{s},\tilde{z}|x}}} \\ &
[(-\log p_{\bm{\tilde{z}}}(\bm{\tilde{z}})- \log p_{\bm{\hat{s}|\tilde{z}}}(\bm{\hat{s}|\tilde{z}}))+ \mathbb{E}_{\bm{y} \sim p_{\bm{y|\hat{s},\tilde{z}}}}[\log p_{\bm{x|y}}(\bm{x|y})]].
\end{split}
\vspace{-2.4em}
\end{equation}
The first term is seen to be identical to the cross entropy between the marginal $q_{\bm{\tilde{z}}}(\bm{\tilde{z}})=\mathbb{E}_{\bm{x} \sim p_{\bm{x}}}\mathbb{E}_{\bm{\hat{s}}\sim q_{\bm{\hat{s}|x}}}[q_{\bm{\hat{s},\tilde{z}|x}}(\bm{\hat{s},\tilde{z}|x})]$ and the prior $p_{\bm{\tilde{z}}}(\bm{\tilde{z}})$, it represents the cost of encoding the side information assuming $p_{\bm{\tilde{z}}}$ as the entropy model. Note $p_{\bm{\tilde{z}}}(\bm{\tilde{z}})$ is modeled as non-parametric fully factorized density \cite{balle2018variational} as
\vspace{-0.5em}
\begin{equation}
p_{\tilde{\bm{z}}}(\tilde{\bm{z}})= \prod_{i}\left(p_{z_{i}|\bm{\psi}^{(i)}}(\bm{\psi}^{(i)}) \ast \mathcal{U}(-\frac{1}{2},\frac{1}{2})  \right),
\vspace{-0.8em}
\end{equation}
where $\bm{\psi}^{(i)}$ encapsulates all the parameters of $p_{z_{i}|\bm{\psi}^{(i)}}$. In addition, to allow optimization via gradient descent during model training phase \cite{balle2016end}, $\bm{z}$ is added an uniform offset $\bm{o}$ instead of scalar quantization $\bar{\bm{z}} = \lfloor \bm{z} \rceil$, i.e. $ \bm{\tilde{z}} = \bm{z} + \bm{o}$ with $ o_{i} \sim \mathcal{U}\left(-\frac{1}{2},\frac{1}{2}  \right)$.

The second term is similar to the first term, representing the cost of encoding $\bm{\hat{s}}$ and denoting the transmission rate of the speech signal. In practice, we utilize the intermediate proxy variable $\bm{y}$ to derive $p_{\bm{\hat{s}|\tilde{z}}}$. During the transmission process, $\bm{y}$ is directly fed into the JSCC encoder and channel without quantization. The process can be described as
\begin{equation}
p_{\bm{\hat{s}|\tilde{z}}} = W(p_{\bm{s|\tilde{z}}}|\bm{h}) = W(f_{e}(p_{\bm{y|\tilde{z}}})|\bm{h}).
\end{equation}
Thus, the density $p_{\bm{\hat{s}|\tilde{z}}}$ can be approximated to $p_{\bm{s|\tilde{z}}}$ and transformed to $p_{\bm{y|\tilde{z}}}$. Similarly, $\bm{z}$ is added an uniform offset instead of scalar quantization, the transmission rate is constrained proportionally to $-\log P_{\bm{y|z}}(\bm{y|z})$.

The third term represents the log-likelihood of recovering $\bm{x}$. Based on the above analysis, the RD function can be simplified to
%\vspace{-0.3em}
\begin{equation}
\vspace{-0.3em}
L_{\mathrm{RD}}= \mathbb{E}_{\bm{x} \sim p_{\bm{x}}}[-\eta_{y} \log p_{\bm{y|z}}(\bm{y|z})-\eta_{z} \log p_{\bm{\tilde{z}}}(\bm{\tilde{z}})+ \lambda d(\bm{x,\hat{x}})],
\end{equation}
where the Lagrange multiplier $\lambda$ determines the tradeoff between the wireless transmission total bandwidth cost $R$ and the end-to-end distortion $D$. $\eta_{y}$ and $\eta_{z}$ are two hyperparameters to balance channel bandwidth consumption of $\bm{y}$ and $\bm{z}$. $d(\cdot,\cdot)$ is the distortion between the original signals and reconstructed signals.

In our paper, we employ the distortion function $d$ both in time domains and frequency domains. Firstly, we utilize mean square error (MSE) between the original signals $\bm{x}$ and reconstructed signals $\bm{\hat{x}}$ to evaluate the reconstructed error in time domain. Then we compute MFCCs \cite{muda2010voice} for the source signal and reconstructed signal and employ $\mathit{l}_{2}$ distance between MFCC vectors in frequency domains to pursue better human perceptual quality \cite{blau2018perception}.
Considering the spectral tilt phenomenon \cite{chen1995adaptive}, if we use MSE as a loss function in frequency domains, the relative error of high-frequencies will be more significant, leading to the loss of high frequency information. Hence, we finally apply normalized mean squared error (NMSE) as the loss function, which helps the model be optimized toward reconstructing high frequencies and gives speech higher human perceptual quality.

%\vspace{-0.2cm}
\subsection{Modular Implementation Details}
In this part, we will share some details of DSST. We take speech signals as a example. After preprocessing, the input frame $\bm{x}$ is modeled as a vector $\bm{x}\in\mathbb{R}^{B\times C\times L}$, where $B$ is the number of frames, $C$ is the number of sound channels, and $L$ is the frame length.

\begin{figure}[t]
\centering
\vspace{-1em}
\includegraphics[scale=0.54]{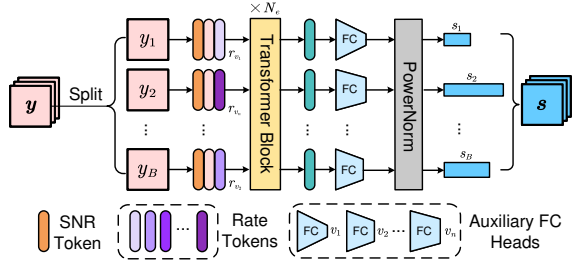}   %0.53
\vspace{-1em}
\caption{The architecture of versatile Deep JSCC encoder. The structure of versatile Deep JSCC decoder is with a mirrored architecture.}
\label{fig3}
\vspace{-1.6em}
\end{figure}

%$\bm{y}^{\prime}\in\mathbb{R}^{\frac{L\times M}{4}\times B}$

\emph{(1) Semantic Analysis and Synthesis Transform $g_{a},g_{s}$}:
The semantic analysis and synthesis transform are used a similar structure as the Kankanahalli network \cite{kankanahalli2018end} , which contains multilayer 1D convolutional blocks and residual networks to extract the speech semantic features efficiently. The latent feature can be written as $\bm{y}\in\mathbb{R}^{B\times M\times \frac{L}{4}}$ after twice downsampling, $M$ is channel dimension.

\emph{(2) Prior Codec $h_{a},h_{s}$}:
The structure of $h_{a}$ and $h_{s}$ is shown at the bottom of of  Fig. \ref{fig2}. It consists of fully convolutional layers followed by the ReLU activation functions. The hyperprior model summarizes the distribution of means and standard deviation in $\bm{z}\in\mathbb{R}^{B\times N\times \frac{L}{16}}$, where $N$ is channel dimension. It effectively captures the semantic temporal connections of the speech waveform.

\emph{(3) Versatile Deep JSCC Codec $f_{e},f_{d}$}:
The architecture of versatile Deep JSCC codec is illustrated in Fig. \ref{fig3}. The encoder consists of $N_{e}$ Transformer blocks and FC layers to implement rate allocation. Specifically, $\bm{y}$ is firstly separated into patch embedding sequence $\left\{y_{1},y_{2},...,y_{B}\right\}$. To adaptively map $y_{i}$ to a $\bar{k}_{i}$-dimensional channel-input vector $s_{i}$, we develop a rate token vector set $\mathcal{R}=\left\{r_{v_{1}},r_{v_{2}},...,r_{v_{n}} \right\}$ to represent rate information and quantization value set $\mathcal{V}=\left\{v_{1},v_{2},...,v_{n} \right\}$ to denote output dimension. By this approach, each $y_{i}$ is merged with its corresponding rate token $r_{v_{i}}$ according to the entropy $-\log P_{y_{i}|\bm{z}}(y_{i}|\bm{z})$ and is fed into the Transformer block and FC layer to transform into a $\bar{k}_{i}$-dimensional vector. In particular, we employ a group of FC layers with different output dimensions $\left\{v_{1},v_{2},...,v_{n} \right\}$, it guided by the rate information accordingly.

Moreover, we perform SNR adaptation in this process. This paper assumes that the transmitter and receiver can receive the SNR feedback information for better performance. As illustrated in Fig. \ref{fig3}, each patch $y_{i}$ is concatenated with SNR token $C\in\mathbb{R}^{B}$, such that the Transformer can learn the SNR information. Hence, the single model can perform at least as well as the models trained for each SNR individually finally when trained under random SNRs.

\emph{(4) Optional Transmission Link}:
If we transmit $\bm{z}$ to obtain the decoding gain, we will quantify $\bm{z}$ first and perform entropy coding and channel coding on it. At the receiver, the prior decoder reconstructs side information $\overline{\bm{\mu}}$, $\overline{\bm{\sigma}}$ after channel decoding and entropy decoding, then feeds them into the Deep JSCC decoder. Accordingly, the total channel bandwidth cost $K$ will increase to $K = K_{y} + K_{z}$, where $K_{z}$ is the bandwidth cost of transmitting $\bm{z}$.

%Our training process takes place in three stages. First, we train $g_{a}$,$g_{s}$,$h_{a}$,$h_{s}$ separately by using allied RD function as loss function to pursue higher compression performance:
%\mathbb{E}_{\bm{\tilde{y}},\bm{\tilde{z}} \sim q_{\bm{\tilde{y},\tilde{z}|x}}}
%\begin{equation}
%L_{RD} = \mathbb{E}_{\bm{x} \sim p_{\bm{x}}}
%[(-logp_{\bm{\tilde{y}|\tilde{z}}}(\bm{\tilde{y}|\tilde{z}}) - logp_{\bm{\tilde{z}}}(\bm{\tilde{z}})) + \lambda d(\bm{x},\bm{\hat{x}})].
%\end{equation}
% Then, we set the gradients of these modules to zero, and train the JSCC codec separately by calculating MSE between the latent feature $\bm{y}$ and reconstructed feature $\bm{\hat{y}}$ as loss function. At last, we turn down the learning rate and fine-tune the model end-to-end by using (5) to trade-off RD performance.
%The goal of optimization of NTC speech coding is to balance the compression ratio and the speech reconstruction quality. Compared to previous residual coding methods, the proposed plain NTC and residual NTC speech coding is trained end-to-end for one shot.

\vspace{-0.5em}
\section{Experiments}
\label{sec:experiments}
\vspace{-0.5em}
In this section, we provide illustrative numerical results to evaluate the quality of speech waveform transmission. Objective quality metrics and a result of a subjective listening test are presented to validate our designed approach.

\subsection{Experimental Setups and Evaluation Metrics}
The waveform is sampled at 16kHz from TIMIT dataset \cite{garofolo1993timit0}.
%which contains 3.1 hours of speech from 462 speakers while the test set contains 0.8 hours of speech. Each speech frame has $L=512$ samples with an overlap of 32 samples.
During total training process, we use Adam optimizer \cite{kingma2014adam} with $B=100$ frames in a single batch and $L=512$ samples in a frame.

To evaluate our model, we use an objective evaluation metric and conduct a subjective experiment as well. Concerning objective quality evaluation, we report perceptual evaluation of speech quality (PESQ) \cite{PESQ} scores to reflect the reconstructed speech quality. Furthermore, we implement MUSHRA subjective test \cite{series2014method} to evaluate audio quality by human raters. We randomly select ten reconstructed waveform signals from the test dataset. To demonstrate the superiority of our model, the comparison schemes include the traditional transmission methods: adaptive multi-rate wideband (AMR-WB) + 5G LDPC and Opus + 5G LDPC \cite{bessette2002adaptive, richardson2018design,valin2012definition}, a standard neural JSCC model DeepSC-S. For traditional separation methods, we choose different coding and modulation schemes according to the adaptive modulation coding (AMC) principle \cite{peng2007adaptive}.

\begin{figure*}[t]
    \centering
    \subfloat[]{
           \includegraphics[width=0.45\columnwidth]{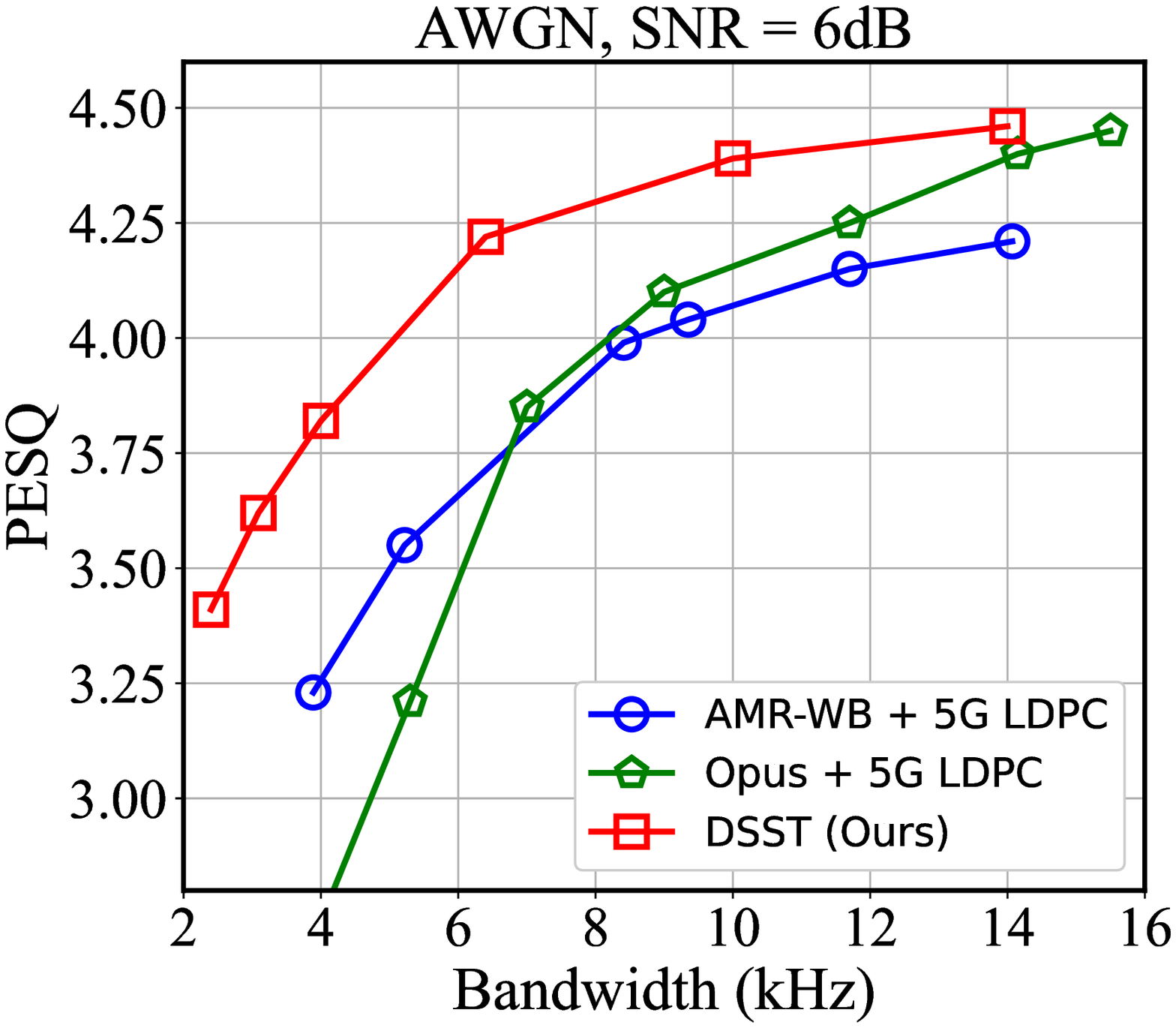}
           \label{fig4a}}
           \hspace{4mm}
    \subfloat[]{
           \includegraphics[width=0.45\columnwidth]{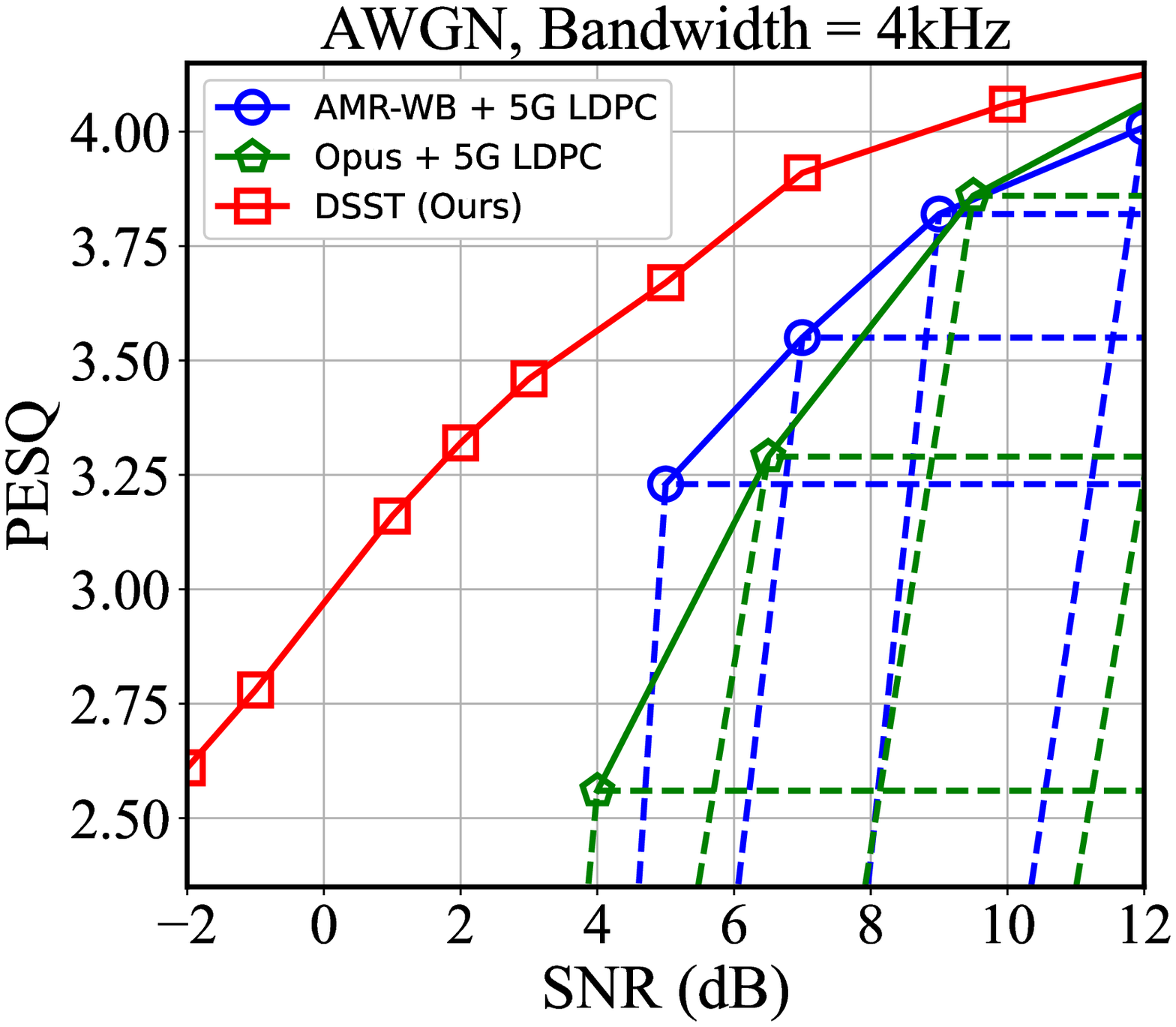}
           \label{fig4b}}
           \hspace{4mm}
    \subfloat[]{
           \includegraphics[width=0.45\columnwidth]{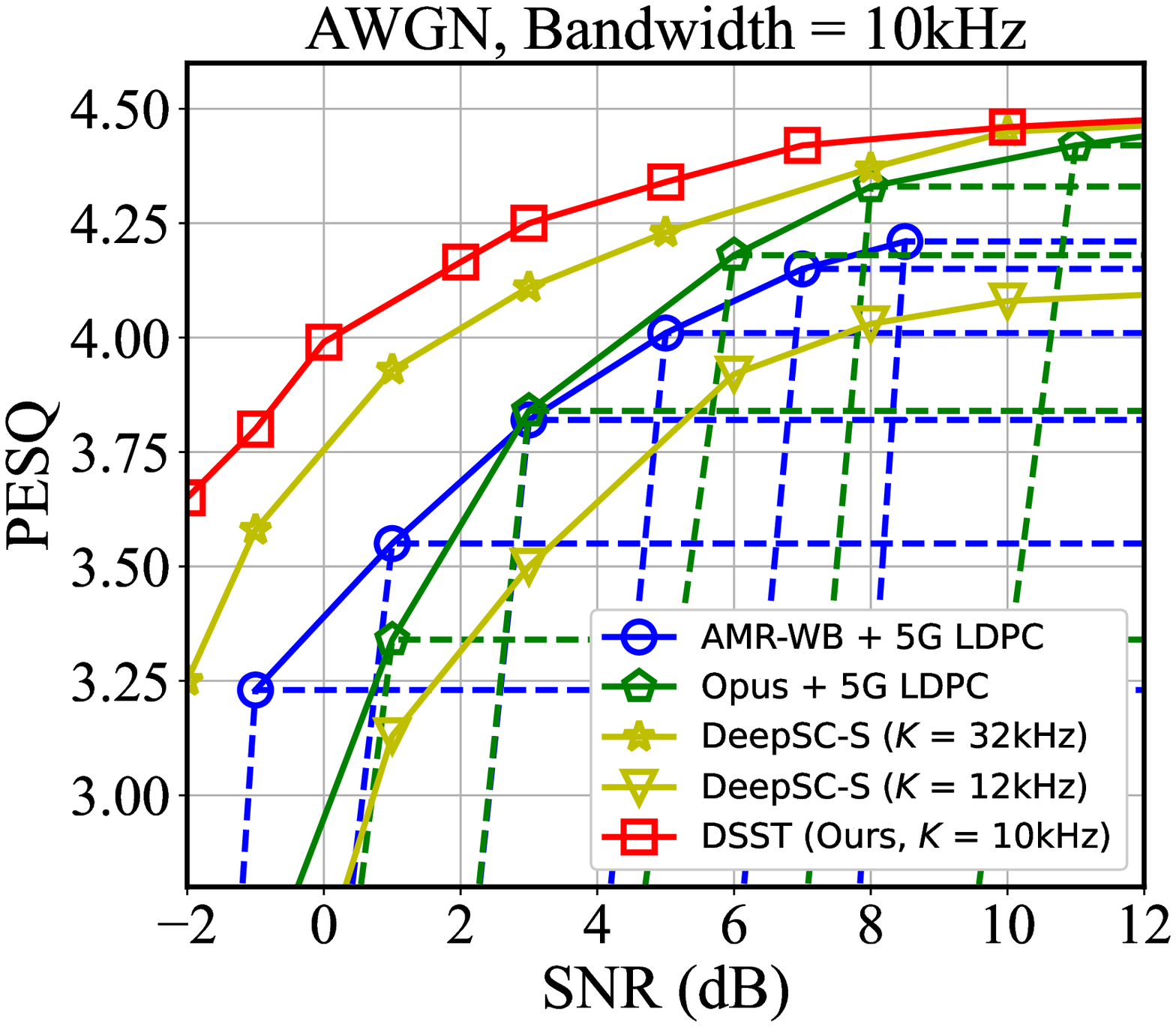}
           \label{fig4c}}
           \hspace{4mm}
    \subfloat[]{
           \includegraphics[width=0.45\columnwidth]{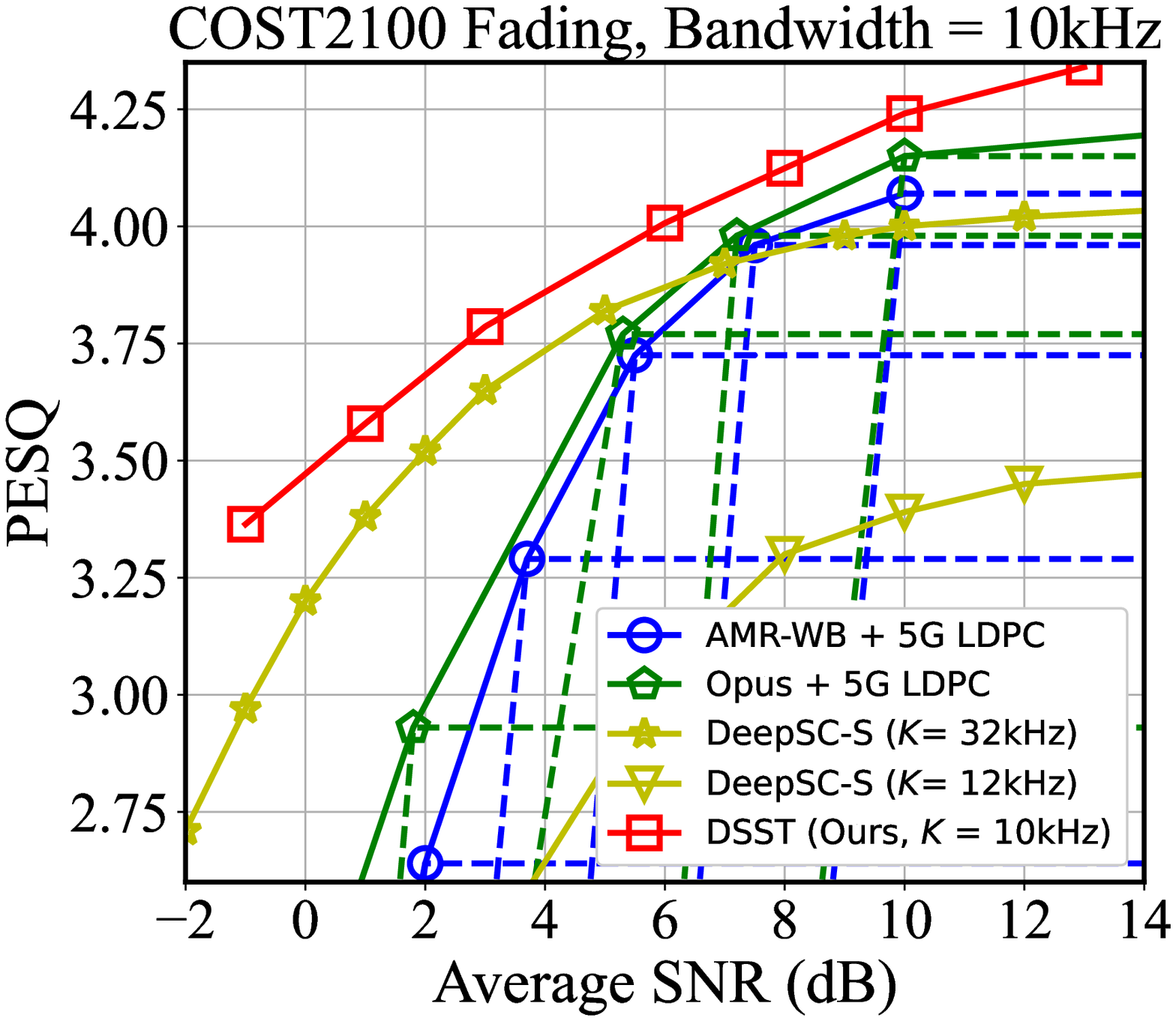}
           \label{fig4d}}
    \vspace{-0.9em}
    \caption{(a) PESQ performance versus the channel bandwidth cost over AWGN channel at SNR = 6dB. (b) PESQ performance versus channel SNR over AWGN channel. The bandwidth cost ($K$) is 4kHz. (c) (d) PESQ performance versus channel SNR over AWGN channel and COST2100 5.3GHz indoor fading channel. $K$ is 10kHz, except DeepSC-S is 12kHz and 32kHz (yellow lines).}
	  \label{fig4}
        \vspace{-0.8em}
\end{figure*}

\begin{figure*}[t]
    \centering
    \subfloat[]{
           \includegraphics[width=0.45\columnwidth]{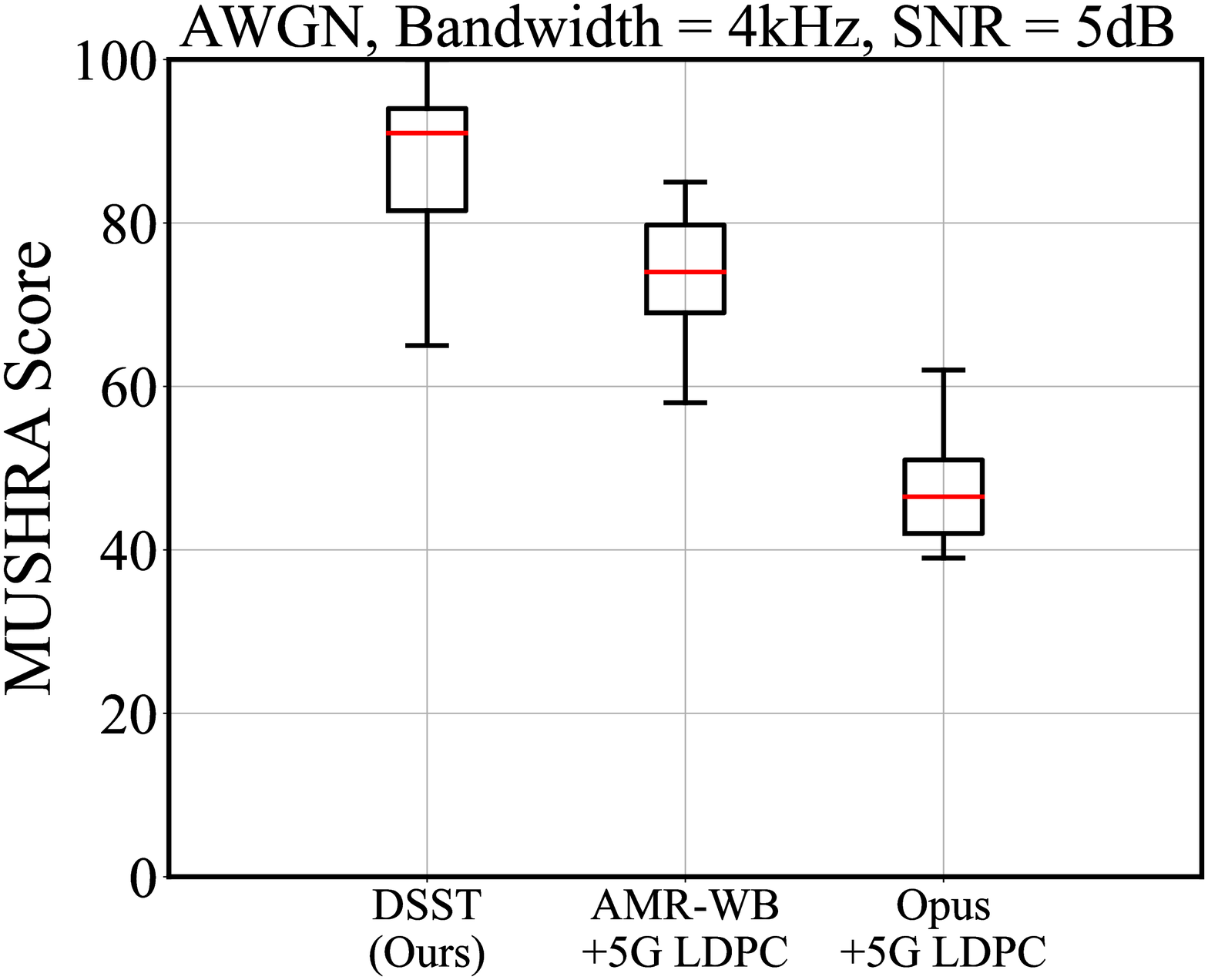}
           \label{fig5a}}
           \hspace{4mm}
    \subfloat[]{
           \includegraphics[width=0.45\columnwidth]{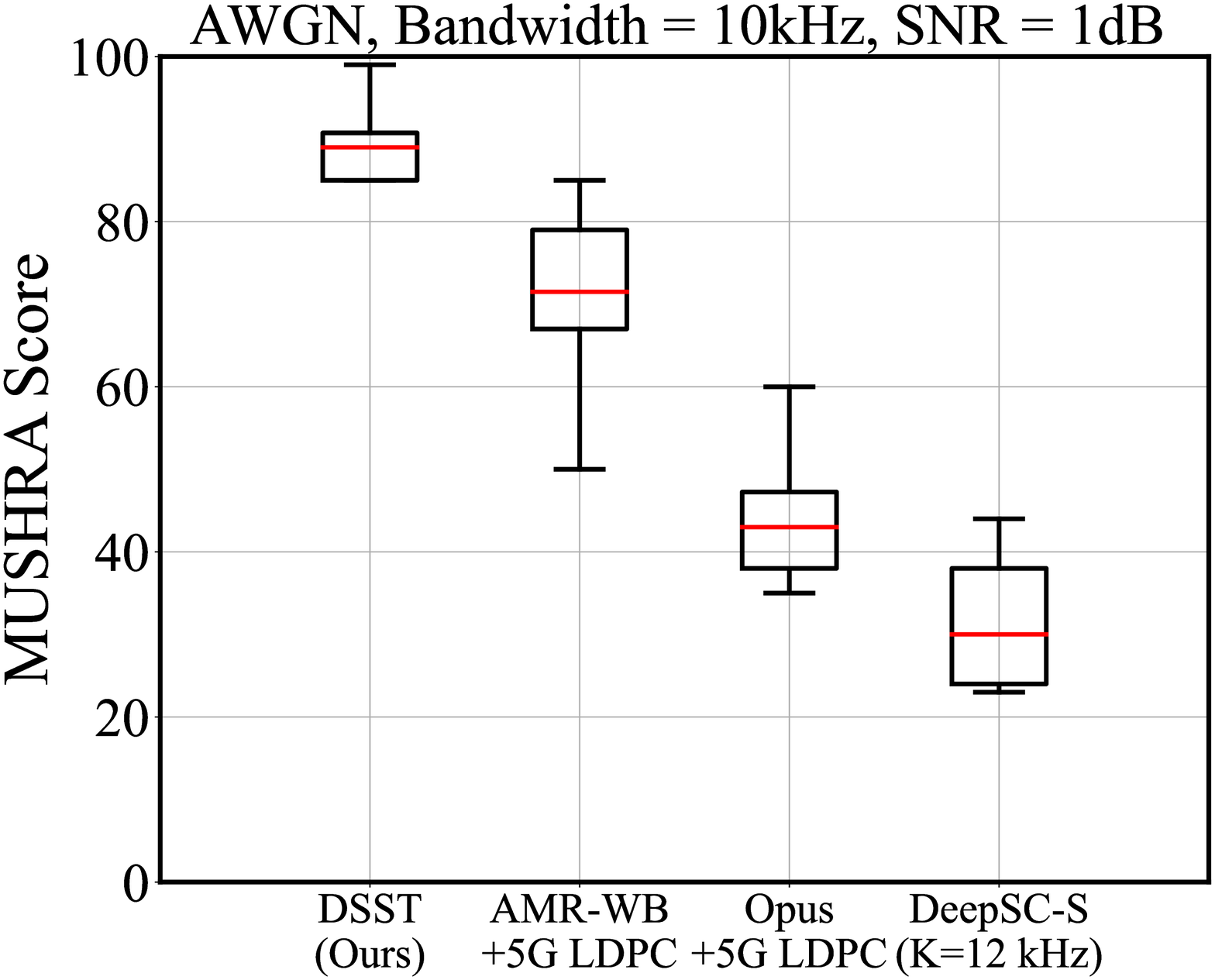}
           \label{fig5b}}
           \hspace{4mm}
    \subfloat[]{
           \includegraphics[width=0.45\columnwidth]{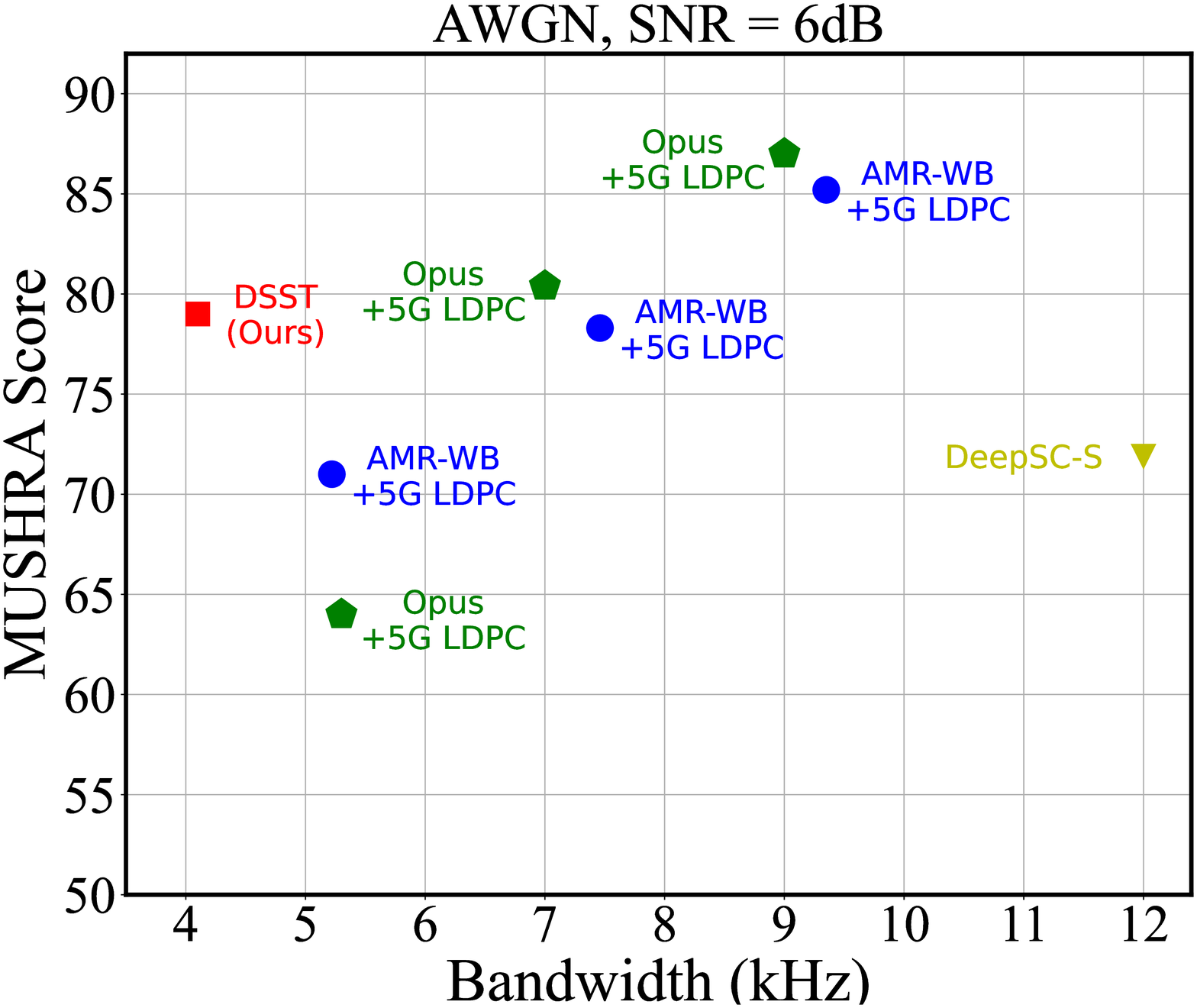}
           \label{fig5c}}
           \hspace{4mm}
        \subfloat[]{
           \includegraphics[width=0.45\columnwidth]{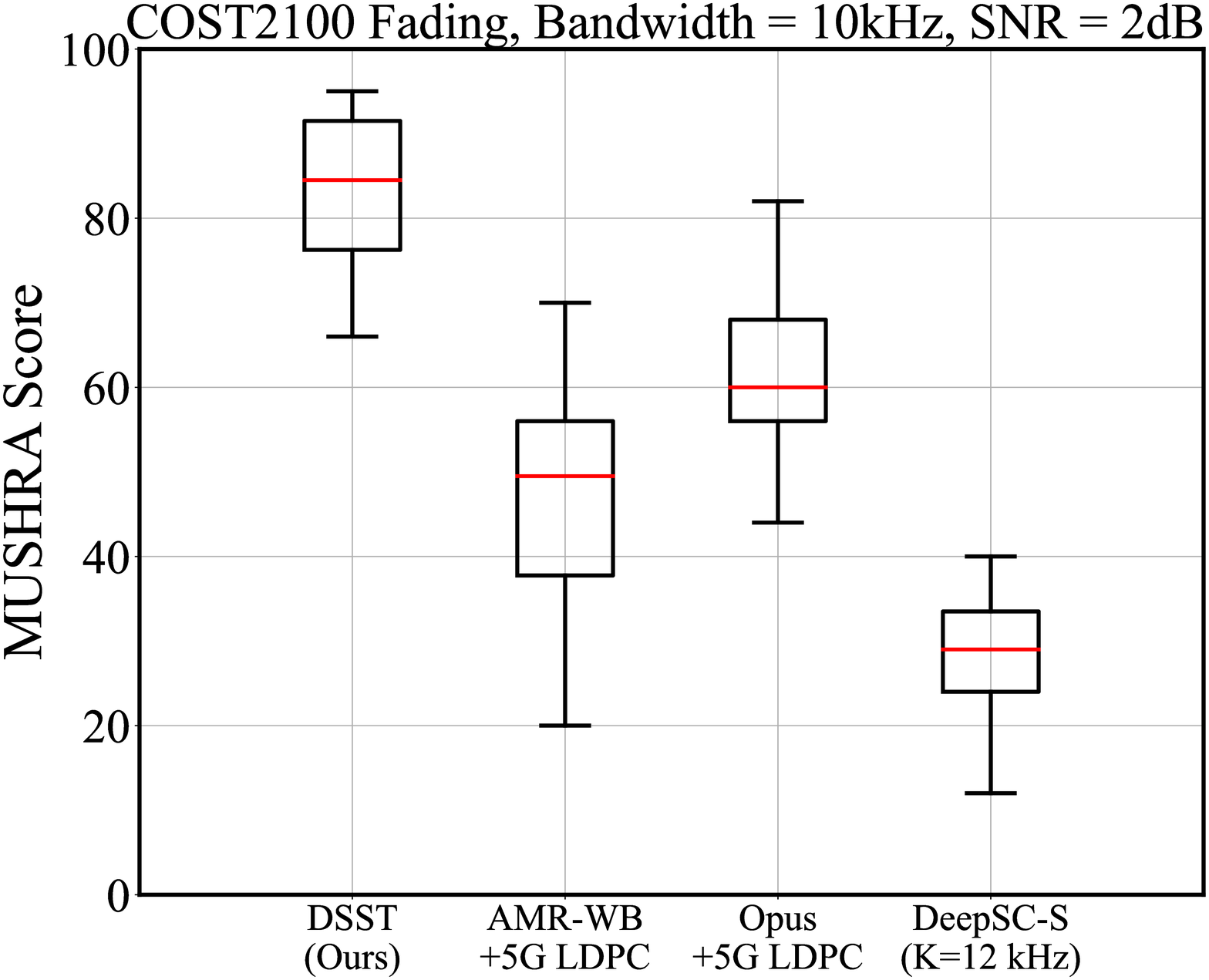}
           \label{fig5d}}
    \label{fig5}
    \vspace{-0.8em}
    \caption{MUSHRA scores for (a) $K$ = 4kHz, SNR = 5dB, AWGN channel. (b) $K$ = 10kHz, SNR = 1dB, AWGN channel. (c) DSST @4kHz vs. classic separation-based schemes and DeepSC-S under AWGN Channel. (d) $K$ = 10kHz, SNR = 5dB, COST2100 fading Channel.}
    \vspace{-1.2em}
\end{figure*}

\subsection{PESQ Performance}
Fig. \ref{fig4}(a) shows the rate-quality results among various transmission methods over the AWGN channel with channel SNR at 6dB. The proposed DSST method outperforms traditional AMR-WB + 5G LDPC and Opus + 5G LDPC under all bandwidth consumption situations. It is mainly because our scheme estimates the distribution of speech source accurately, which gets the compression gain and it mitigates the mismatch between source coding and channel coding compared with classical schemes. Moreover, the DSST scheme demonstrates its flexibility. The model can implement an efficient speech transmission under any transmission rate. Compared with traditional schemes, neither AMR-WB nor Opus schemes can effectively transmit the speech at low bandwidth.

Fig. \ref{fig4}(b) and Fig. \ref{fig4}(c) discuss the performance of the model under different SNRs. Similar to the above findings, the proposed DSST scheme outperforms separation-based traditional schemes by a large margin for all SNRs.
Furthermore, illustrated in Fig. \ref{fig4}(a) and Fig. \ref{fig4}(c), when PESQ is 4.3 and the SNR is 6dB, DeepSC-S uses 32kHz bandwidth, while our model only uses 8kHz bandwidth. It saves about 75$\%$ of channel bandwidth consumption. This gain comes from both the semantic transform network and the versatile Deep JSCC network. The former drops most of the redundancy in speech signals. The latter allocates different coding rates more reasonably, which gets the most out of coding resources.
%maximizes the system coding gain.

To further verify the effectiveness of our model, we carry out experiments on the practical COST2100 fading channel. CSI samples are collected in an indoor scenario at 5.3GHz bands, and all schemes use one-shot transmission. Illustrated in Fig. \ref{fig4}(d), DSST shows strong robustness to channel variations. Especially under the condition of low SNRs, our model incomparably outperforms other schemes. Compared with existing neural methods DeepSC-S, our model achieves more than a 30$\%$ increase in terms of PESQ performance when they use similar bandwidth resources.

\subsection{User Study}
To better align with human perception, we have conducted the MUSHRA subjective test user study. The first three figures show the results under the AWGN channel. Fig. 5(a) shows the performance under the low bandwidth cost condition. The DSST significantly surpasses two separate coding schemes. Fig. 5(b) reveals the performance under a low SNR situation. We observe similar results, and our model achieves a three times higher MUSHRA score than DeepSC-S with similar bandwidth. Fig. 5(c) compares various transmission schemes with the DSST under different bandwidth costs. To match the speech quality of DSST, Opus + 5G LDPC needs to use 7kHz bandwidth, while AMR-WB + 5G LDPC needs at least 7.5kHz bandwidth. It saves more than 45$\%$ of bandwidth resources. Moreover, DSST reconstructed high-quality speech using as little as 4kHz bandwidth, with quality significantly better than DeepSC-S at 12kHz. At last, Fig. 5(d) illustrates the performance over the COST2100 fading channel. Our model still achieves better performance and outperforms other transmission schemes.

We provide examples of reconstructed speech for an auditory comparison at
\emph{https://ximoo123.github.io/DSST}.

%The experiment is mainly conducted under two harsh channel conditions: low bandwidth cost and low SNR.
\section{Conclusion}
%\vspace{-0.1cm}
\label{sec:conclusion}
We present the DSST, a new class of high-efficiency semantic speech transmission model over the wireless channel. The model first extracts the semantic features of the speech waveform and then transmits it. Particularly, our model takes advantage of the nonlinear transform method, learnable entropy model, and versatile Deep JSCC method to get higher coding gain. Results indicate that our model achieves better performance under both the AWGN channel and COST2100 fading channel. It is exciting that the DSST reconstructs high-quality speeches under harsh channel conditions. It will be of great help to future speech semantic communication.

\vfill
\pagebreak

\bibliographystyle{IEEEbib}
\bibliography{bliography}

\begin{thebibliography}{10}

\bibitem{shannon1948mathematical}
Claude~Elwood Shannon,
\newblock ``A mathematical theory of communication,''
\newblock {\em The Bell System Technical Journal}, vol. 27, no. 3, pp.
  379--423, 1948.

\bibitem{kurka2021bandwidth}
David~Burth Kurka and Deniz G{\"u}nd{\"u}z,
\newblock ``Bandwidth-agile image transmission with deep joint source-channel
  coding,''
\newblock {\em IEEE Transactions on Wireless Communications}, vol. 20, no. 12,
  pp. 8081--8095, 2021.

\bibitem{farsad2018deep}
Nariman Farsad, Milind Rao, and Andrea Goldsmith,
\newblock ``Deep learning for joint source-channel coding of text,''
\newblock in {\em 2018 IEEE International Conference on Acoustics, Speech, and
  Signal Processing (ICASSP)}. IEEE, 2018, pp. 2326--2330.

\bibitem{Dai}
Jincheng Dai, Sixian Wang, Kailin Tan, Zhongwei Si, Xiaoqi Qin, Kai Niu, and
  Ping Zhang,
\newblock ``Nonlinear transform source-channel coding for semantic
  communications,''
\newblock {\em IEEE Journal on Selected Areas in Communications}, vol. 40, no.
  8, pp. 2300--2316, 2022.

\bibitem{Dai9852388}
Jincheng Dai, Ping Zhang, Kai Niu, Sixian Wang, Zhongwei Si, and Xiaoqi Qin,
\newblock ``Communication beyond transmitting bits: Semantics-guided source and
  channel coding,''
\newblock {\em IEEE Wireless Communications}, pp. 1--8, 2022.

\bibitem{weng2021semantic}
Zhenzi Weng and Zhijin Qin,
\newblock ``Semantic communication systems for speech transmission,''
\newblock {\em IEEE Journal on Selected Areas in Communications}, vol. 39, no.
  8, pp. 2434--2444, 2021.

\bibitem{liu2012cost}
Lingfeng Liu, Claude Oestges, Juho Poutanen, Katsuyuki Haneda, Pertti
  Vainikainen, Fran{\c{c}}ois Quitin, Fredrik Tufvesson, and Philippe
  De~Doncker,
\newblock ``The {COST} 2100 {MIMO} channel model,''
\newblock {\em IEEE Wireless Communications}, vol. 19, no. 6, pp. 92--99, 2012.

\bibitem{balle2018variational}
Johannes Ball{\'e}, David Minnen, Saurabh Singh, Sung~Jin Hwang, and Nick
  Johnston,
\newblock ``Variational image compression with a scale hyperprior,''
\newblock in {\em Proceedings of the International Conference on Learning
  Representations (ICLR)}, 2018.

\bibitem{balle2020nonlinear}
Johannes Ball{\'e}, Philip~A Chou, David Minnen, Saurabh Singh, Nick Johnston,
  Eirikur Agustsson, Sung~Jin Hwang, and George Toderici,
\newblock ``Nonlinear transform coding,''
\newblock {\em IEEE Journal of Selected Topics in Signal Processing}, vol. 15,
  no. 2, pp. 339--353, 2020.

\bibitem{balle2016end}
Johannes Ball{\'e}, Valero Laparra, and Eero~P. Simoncelli,
\newblock ``End-to-end optimized image compression,''
\newblock in {\em Proceedings of the International Conference on Learning
  Representations (ICLR)}, 2017.

\bibitem{muda2010voice}
Lindasalwa Muda, Mumtaj Begam, and Irraivan Elamvazuthi,
\newblock ``Voice recognition algorithms using mel frequency cepstral
  coefficient ({MFCC}) and dynamic time warping ({DTW}) techniques,''
\newblock {\em arXiv preprint arXiv:1003.4083}, 2010.

\bibitem{blau2018perception}
Yochai Blau and Tomer Michaeli,
\newblock ``The perception-distortion tradeoff,''
\newblock in {\em Proceedings of the IEEE Conference on Computer Vision and
  Pattern Recognition (CVPR)}, 2018, pp. 6228--6237.

\bibitem{chen1995adaptive}
Juin-Hwey Chen and Allen Gersho,
\newblock ``Adaptive postfiltering for quality enhancement of coded speech,''
\newblock {\em IEEE Transactions on Speech and Audio Processing}, vol. 3, no.
  1, pp. 59--71, 1995.

\bibitem{kankanahalli2018end}
Srihari Kankanahalli,
\newblock ``End-to-end optimized speech coding with deep neural networks,''
\newblock in {\em 2018 IEEE International Conference on Acoustics, Speech, and
  Signal Processing (ICASSP)}. IEEE, 2018, pp. 2521--2525.

\bibitem{garofolo1993timit0}
John~S Garofolo,
\newblock ``{TIMIT} acoustic phonetic continuous speech corpus,''
\newblock {\em Linguistic Data Consortium}, 1993.

\bibitem{kingma2014adam}
Diederik~P Kingma and Jimmy Ba,
\newblock ``Adam: A method for stochastic optimization,''
\newblock in {\em Proceedings of the International Conference on Learning
  Representations (ICLR)}, 2015.

\bibitem{PESQ}
A.W. Rix, J.G. Beerends, M.P. Hollier, and A.P. Hekstra,
\newblock ``Perceptual evaluation of speech quality ({PESQ})-a new method for
  speech quality assessment of telephone networks and codecs,''
\newblock in {\em 2001 IEEE International Conference on Acoustics, Speech, and
  Signal Processing (ICASSP)}, 2001, vol.~2, pp. 749--752 vol.2.

\bibitem{series2014method}
B~Series,
\newblock ``Method for the subjective assessment of intermediate quality level
  of audio systems,''
\newblock {\em International Telecommunication Union Radiocommunication
  Assembly}, 2014.

\bibitem{bessette2002adaptive}
Bruno Bessette, Redwan Salami, Roch Lefebvre, Milan Jelinek, Jani
  Rotola-Pukkila, Janne Vainio, Hannu Mikkola, and Kari Jarvinen,
\newblock ``The adaptive multirate wideband speech codec ({AMR-WB}),''
\newblock {\em IEEE Transactions on Speech and Audio Processing}, vol. 10, no.
  8, pp. 620--636, 2002.

\bibitem{richardson2018design}
Tom Richardson and Shrinivas Kudekar,
\newblock ``Design of low-density parity check codes for 5{G} new radio,''
\newblock {\em IEEE Communications Magazine}, vol. 56, no. 3, pp. 28--34, 2018.

\bibitem{valin2012definition}
Jean-Marc Valin, Koen Vos, and Timothy Terriberry,
\newblock ``Definition of the {O}pus audio codec,''
\newblock Tech. {R}ep., 2012.

\bibitem{peng2007adaptive}
Fei Peng, Jinyun Zhang, and William~E Ryan,
\newblock ``Adaptive modulation and coding for {IEEE} 802.11 n,''
\newblock in {\em 2007 IEEE Wireless Communications and Networking Conference}.
  IEEE, 2007, pp. 656--661.

\end{thebibliography}

\end{document}